# Tantalum Carbide Nanoparticles as Enzyme Mimics for X-Ray Computed Tomography Imaging and Unlabeled Localization in Mice

Tongming Chen[1†], Xiumei Tian[2†], Xiaoju Wu[1], Ao Zeng[2], Yuan Chen[1], Guowei Yang[1]*

[1] State Key Laboratory of Optoelectronic Materials and Technologies, Nanotechnology Research Center, School of Physics, School of Materials Science & Engineering, Sun Yat-sen University, Guangzhou 510275, Guangdong, P. R. China

[2] Department of Biomedical Engineering, School of Basic Medical Sciences, Affiliated Stomatology Hospital and Sino-French Hoffmann Institute, Guangzhou Medical University, Guangzhou 510182, P. R. China

[†] These authors contributed equally to this work.

* Corresponding author: stsygw@mail.sysu.edu.cn

**Abstract**

Nanoparticles contrast medium to enhance performance of X-ray computed tomography (CT) imaging are desirable. Their clinical translation requires biodistribution and pharmacodynamics assessing in vivo. Herein, we for the first time reported that tantalum carbide nanoparticles (TCNPs) synthesized by a one-step laser ablation method are suitable probes for CT imaging. Measurements showed that Hounsfield units value of TCNPs was greater than that of Iohexol in the same concentration and in vivo X-ray CT imaging using TCNPs accounted for bright and high-resolution CT images with long circulation time. More importantly, we made the surprising discovery that TCNPs catalyze the color reaction of enzyme substrate incorporated with $H_2O_2$, similar to that of natural horseradish peroxidase. Based on this reaction, a simple one-step histochemical method was proposed for visualizing and quantifying the localization of TCNPs in the tissues or organs and further determining their biodistribution in mice. The novel approach avoids complex labeling and renders the possibility to learn the real state of TCNPs as contrast medium. These results proposed TCNPs are promising dual-functional agents for imaging and ex vivo localization.

**Introduction**

Tantalum is a hard, ductile, and highly chemical-resistant metal with excellent bioperformance, developed as stents, sutures, and implants[1]. It is known to be inherently bioinert, demonstrating no remarkable inflammatory response under in vivo conditions. Depending on the required application, tantalum based materials can be usually available in bulk form and in coating form[2]. Recently, tantalum oxide nanoparticles have been proposed as promising agents for X-ray computed tomography (CT) imaging[3]. Present CT contrast medium approved for clinical use are iodine containing molecules[4-5]. They are efficient in absorbing X-rays, but rapidly cleaned by kidneys due to their small size, resulting in short circulation time. Moreover, it is necessary to use high concentration of iodine containing agents to provide adequate contrast, which may cause serious side-effects in patients[6-9]. As compared to conventional iodine containing molecules, tantalum oxide nanoparticles are featured by significant contrast increase and extended blood circulation time[10].

Clinical translation of nanoparticles in CT imaging requires extensive risk assessment for biodistribution and pharmacodynamics profiles after injection are crucial. Nevertheless, precise detection of nanoparticles ex/in vivo remains elusive. Factors such as surface properties, hydrodynamic size, and dispersibility are expected to influence interactions with biological systems[11-13]. Hence, biodistribution studies are often performed employing optical and radioisotopic tracers conjugated to nanoparticles as imaging agents[14-15]. However, the surface properties of nanoparticles and their distribution in body are probably being affected by exogenous labeling. Furthermore, the inevitable deposition of labels often causes false detection results in long-term monitoring[16]. Thus, developing new technology by using their intrinsic characteristics to ensure in situ detection of nanoparticles is desirable. In recent years, ex vivo detection of $Fe_3O_4$ nanoparticles in mice using their enzymatic activity was proposed, showing higher sensitivity than traditional Prussian blue assay[17].

Here we synthesized TCNPs by a one-step laser-induced method, i.e., laser ablation in liquid (LAL)[18-19]. Our experiments showed that Hounsfield units value of TCNPs was greater than that of clinical iodinated molecules in equivalent dose. Furthermore, in vivo X-ray CT imaging using TCNPs accounted for bright and high-resolution CT images with long circulation time. More importantly, we made the surprising discovery that TCNPs possess peroxidase-mimicking activity, catalyzing the oxidation of natural enzyme substrate incorporated with $H_2O_2$ to generate a characteristic color reaction. Based on this reaction, a simple histochemical method was proposed for visualizing and quantifying the localization of TCNPs in the tissues or organs and further determining their biodistribution and pharmacodynamics in mice (Scheme 1). The novel approach avoids complex labeling and renders the possibility to determine the real state of TCNPs as contrast agent ex vivo by using their intrinsic properties.

**Results and discussion**

After laser ablation of Ta target in absolute ethanol for 5 min, color of the suspension changed from colorless to grey. The supernatant was collected for further characterization. TEM image in Figure 1a proposes that the products are mostly of spherical shape. A statistical analysis of 100 particles indicates that the diameter of

TCNPs, ranging from 10-100 nm, have an average size of 40 nm. EDS spectrum demonstrates the main ingredient elements of Ta and C (Figure 1b). As shown in Figure 1c, the corresponding XRD measurements propose the patterns of the laser ablated products are indexed to two forms of tantalum carbide including $Ta_4C_3$ with cubic structure (JCPDS card No. 65-0025) and $Ta_2C$ with trigonal structure (JCPDS card No. 65-3191). HRTEM image in Figure 1d further supports the results of XRD patterns, demonstrating clear lattice fringes with planer spacing of 0.236 nm, 0.247 nm, and 0.221 nm, in good agreement with (011) and (002) crystal plane of $Ta_2C$, and (200) crystal plane of $Ta_4C_3$, respectively. The surface chemical states of Ta in TCNPs were analyzed by XPS spectrum. The high resolution spectrum of the Ta 4f peak could be decomposed into six main components (Figure 1e)[20]. Two high energy peaks at 25.90 and 27.88 eV corresponding to $4f_{7/2}$ and $4f_{5/2}$ components of $Ta^{5+}$ and four low energy peaks at 23.61 and 22.89 eV assigned to the $4f_{7/2}$ peaks of $Ta^{3+}$ and $Ta^{2+}$, respectively, and at 26.88 and 24.59 eV assigned to the $4f_{5/2}$ components of $Ta^{3+}$ and $Ta^{2+}$, respectively. The $Ta^{3+}$ and $Ta^{2+}$ signals can be derived from $Ta_4C_3$ and $Ta_2C$ phases in TCNPs, respectively, and the oxidized (5+) state of Ta is probably deriving from the defects or oxides formed on the surface of TCNPs. ICP-AES test stressed that the TCNPs is about 74% by weight tantalum. The particle size distribution of the TCNPs solution was monitored by DLS measurements. The peak of the particle hydrodynamic diameters distribution locates at around 378 nm (Figure 1f).

Biocompatibility assays are needed for application of TCNPs in vitro and in vivo. The in vitro experiment of the TCNPs was examined by CCK-8 assay[21]. GES-1 cells and HEp-2 cells viability was recorded at 12 h, 24 h, and 36 h after incubation with varied concentrations (1−20 μM Ta) of TCNPs, respectively. Control experiments were performed in parallel without TCNPs. As shown in Figure 1g and h, TCNPs did not significantly affect GES-1 cells viability up to 20 μM Ta. Further, in the case of HEp-2 cells, there were not distinction between negative control group and TCNPs groups with time and concentration. Hence, these results propose that TCNPs have insignificant harmful effects on cells viability and differentiation, and the cytotoxicity of the TCNPs is relative satisfactory.

Though the TCNPs with minimal cytotoxcity and good efficiency *in vitro*, it is essential for nano-based biomedicine to evaluate their immunotoxicity *in vivo*. In general, the expression of clusters of differentiation (CD) are very crucial for the immune response. For example, CD206 and CD11b are CD markers of innate immunity and CD25, CD69 and CD71 are are CD markers of adaptive immunity[22-24]. Their expression levels are an important parameter in evaluating the immune response[25]. The results of immunotoxicity on the TCNPs are shown in Figure 2. As we can see that, there are insignificant diffterences for expression levels of all the CD markers between TCNPs group and Gd-DTPA group at 7d after injection (P=0.05). It is demonstrated that the TCNPs do not stimulate the innate and adaptive cells after injection, and the immunotoxicity *in vivo* of TCNPs is also satisfactory.

To assess the effect of CT contrasts, the X-ray absorption of TCNPs was compared to that of Iohexol in vitro. Figure 3a shows a high positive correlation between the Housfield units (HU) values and concentration for both TCNPs and Iohexol. At equivalent concentration, the HU value of TCNPs was greater than that of Iohexol since the attenuation coefficient of Ta (4.3 cm$^2$ kg$^{-1}$ at 100 KeV) was higher than that of I (1.94 cm$^2$ kg$^{-1}$ at 100 KeV)[3, 10]. The X-ray CT phantom images in Figure 3b show that TCNPs offered significantly higher contrast enhancement than Iohexol (Figure 3b). The important problem with the development of CT contrast agents is their relatively low sensitivity of the CT signal[8]. The TCNPs developed by us demonstrate an increased contrast at an equivalent dose in comparison with the clinical Iodinated molecules. Hence, TCNPs as contrast agents are able to show CT imaging signal with higher sensitivity. Further, lower dose requirement in this case may benifit a contrast agent because it decreases underlying side-effects in body in clinical applications[7]. In view of its low cytotoxicity and high CT efficacy, we then evaluated the feasibility of TCNPs as a CT contrast agent in vivo. TCNPs suspended in saline solution was intratumorally injected into a mouse. The CT imaging were performed before injection as well as 1h injection, and side-effects were examined in the tumor. Figure 3c shows that, after administration of TCNPs, clear contrast enhancement than other soft tissue could be observed at injection site, because of the TCNPs inducing a strong X-ray attenuation. Additionally, 3D-renderings of CT images were performed. The results also present distinct improvement of the tumor signals at the same time intervals (Figure 3d).

To study the nanozyme activity of TCNPs, color reaction was performed in the absence and presence of H$_2$O$_2$, choosing TMB as the substrate[26-28]. With the involvement of H$_2$O$_2$, the TMB color evolution was much faster and exhibited deeper color in the same period of time (Figure 4a). However, the control experiments without

TCNPs showed negligible color variation, proposing that all these components are necessary to stimulate the reaction. The UV-vis absorbance spectrum of the blue product showed absorbance peak at 652 nm, exhibiting the POD-like activity of TCNPs (Figure 4b)[29-32]. The origin of the POD-like activity should also be considered worthy of attention. The Ta 4f XPS spectrum in Figure 1e proposes the presence of $Ta^{3+}$, $Ta^{2+}$, and defects on the surface of TCNPs. We proposed that the $Ta^{3+}/Ta^{2+}$ redox couple can promote the electron transfer process between $H_2O_2$ and TMB in a $Ta^{3+} \leftrightarrow Ta^{2+}$ switching manner, which is responsible for the peroxidase mimetic behavior[29, 33]. Also, the defects are also likely to take a role in concerning the origin of the POD-like activity[34-35]. Further experimental evidences are needed.

By taking advantage of the POD-like activity of TCNPs that can produce a color precipitate, we first evaluated the cell permeability of the TCNPs. The TCNPs and TMB incubated HEp-2 cells show a distinct intracellular blue color deposition in the microscope image, indicating that the TCNPs can easily permeate into the cell (Figure 4c). The DAB incubated cells show similar results, forming a brown color deposition at the site of TCNPs (Figure 4d). These results propose the feasibility of this location method and the possibility of detection the biodistribution of TCNPs ex vivo[17]. After intravenously administrated by TCNPs, the mice were sacrificed and their organs were collected and embedded in paraffin sections and stained with POD-like assay using TMB as substrate. The TCNPs can catalyze the oxidation of TMB in the presence of $H_2O_2$ and make a blue deposition at the sites of TCNPs (Figure 5). We can see that TCNPs were localized clearly in heart, lung, and liver. Rare intake of TCNPs was observed in the other organs. Hence, visualizing TCNPs in tissues by utilizing their peroxidase activity was realized, avoiding exogenous labeling. Further, the pharmacokinetics and organ clearance of CT contrast agents can be factually evaluated to understand their in vivo behavior.

**Conclusion**

In summary, we synthesized TCNPs by a one-step method of laser ablation in liquid. Our measurements indicated that TCNPs show higher HU value than that of clinical iodinated molecules at equivalent concentration. Furthermore, in vivo CT imaging using TCNPs resulted in bright and high-resolution CT images. More importantly, we found that TCNPs possess peroxidase-mimicking activity, catalyzing the oxidation of natural enzyme substrate in the presence of $H_2O_2$ to produce a characteristic color reaction. Based on this reaction, a histochemical method was proposed for visualizing and quantifying the localization of TCNPs in the main tissues or organs and further determining their biodistribution in mice. The novel approach avoids complex labeling and offers the possibility to learn the real state of TCNPs as contrast agent ex vivo by using their intrinsic enzyme mimetic properties.

**Conflicts of interest**

The authors declare there are no conflicts of interest.

**Acknowledgement**

The National Basic Research Program of China (2014CB931700), the National Science

Foundation for Young Scholars of China (81401462), the National Science Foundation of China (81771891), Science and Technology Project of Guangdong Province (2017B090911012), and State Key Laboratory of Optoelectronic Materials and Technologies supported this work.

**Figure Captions**

**Scheme 1.** Schematic illustration of the preparation of TCNPs by LAL and their applications for CT imaging and unlabeled localization.

**Figure 1. Characterizations of TCNPs.** (a) TEM image of TCNPs. The inset is the statistic analysis of diameter of TCNPs. (b) EDS spectrum of TCNPs. (c) XRD spectrum of TCNPs. (d) HRTEM image of TCNPs. (e) High-resolution Ta 4f XPS spectra of TCNPs. (f) Size distribution of the TCNPs obtained by DLS analysis. (g) Cytotoxicity of TCNPs on HEp-2 cells as function of the concentration. (h) Cytotoxicity of TCNPs on GES-1 and HEp-2 cells with varied incubation time.

**Figure 2.** *In vivo* **immunotoxicity of TCNPs.** (a) Flow cytometry scatter plot of monocytes and lymphocytes group in peripheral blood after injection TCNPs in Balb/c mice. (1) PE-CD3, FITC-CD25, (2) PE-CD3, FITC-CD69, (3) PE-CD3, FITC-CD71, (4) APC-F4/80, PE-CD206, (5) APC-F4/80, PE-CD11b. (b) Mean fluorescence intensity (MFI) of different CD markers in the monocytes and lymphocytes of peripheral blood at 7d after the injection of TCNPs by the tail veins of BALB/c mice (100 mM Ta, mean ± SD, n = 5 in each group).

**Figure 3. In vitro and in vivo CT imaging.** (a) HU values of TCNPs and Iohexol with different concentrations. The error bars represent the standard deviation of three measurements. (b) In vitro CT images of TCNPs and Iohexol with different

concentrations. (c) Serial CT coronal views of in vivo imaging on a mouse following intratumoral injection of TCNPs. (d) The corresponding 3D renderings of in vivo CT images.

**Figure 4.** (a and b) TCNPs show peroxidase-like activity. The TCNPs catalyze the oxidation of TMB, in the presence of $H_2O_2$. (1) TCNPs + TMB, (2) TMB + $H_2O_2$, (3) TCNPs + TMB + $H_2O_2$. TCNPs-peroxidase staining of HEp-2 cells using substrates of (c) TMB and (d) DAB, respectively. Arrows indicate some of the stained TCNPs. The insets show the high resolution images of HEp-2 cells.

**Figure 5.** TCNPs-peroxidase staining of paraffin-embedded heart, liver, and lung from TCNPs-administered mice. Arrows indicate some of the stained TCNPs.

**Scheme 1**

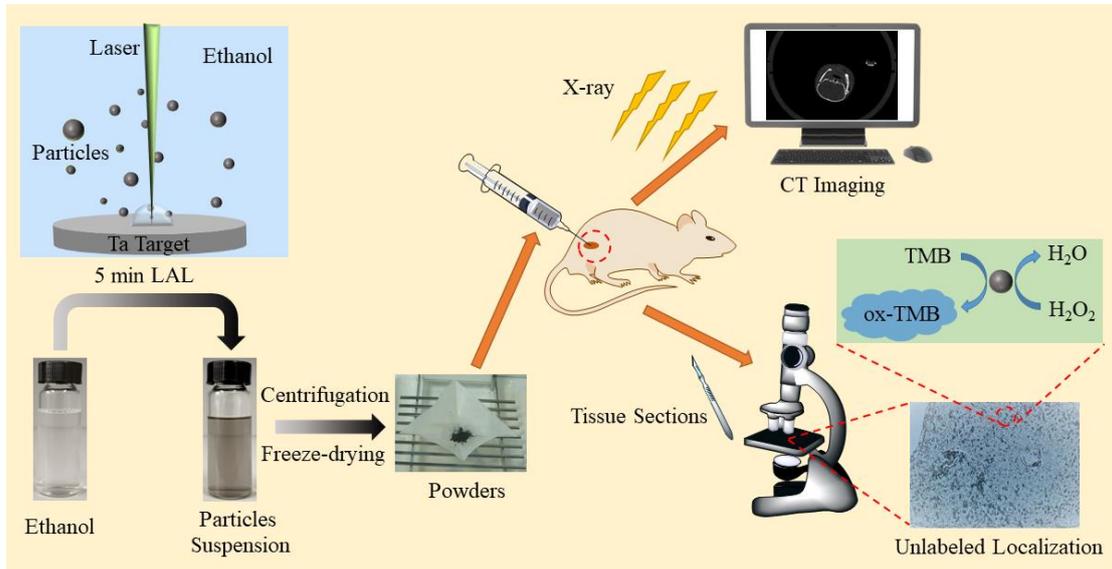

**Figure 1**

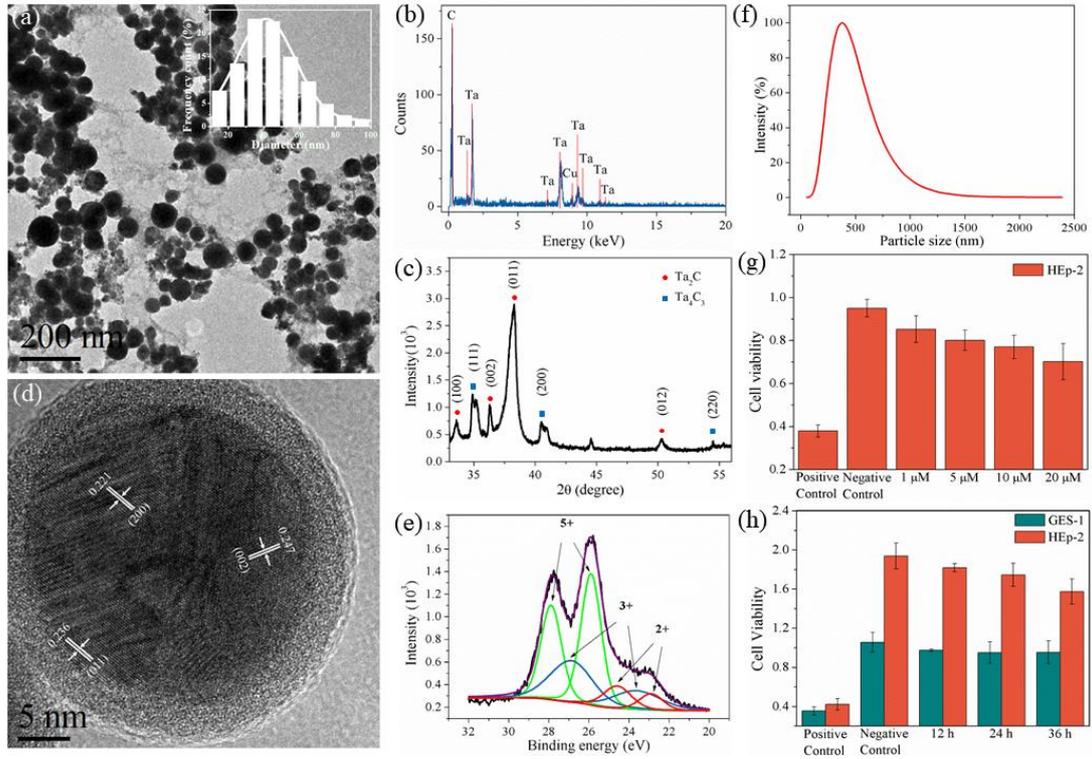

**Figure 2**

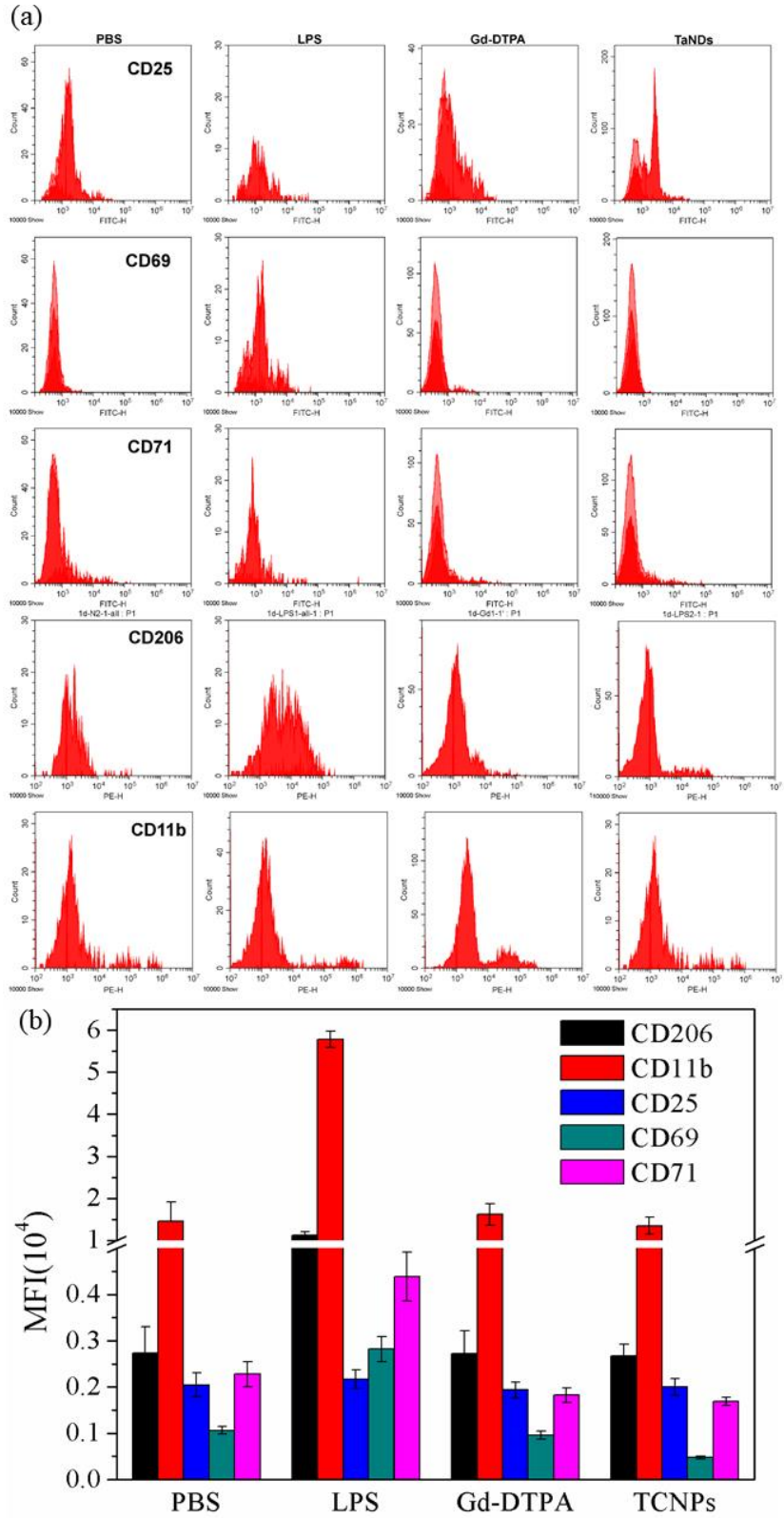

**Figure 3**

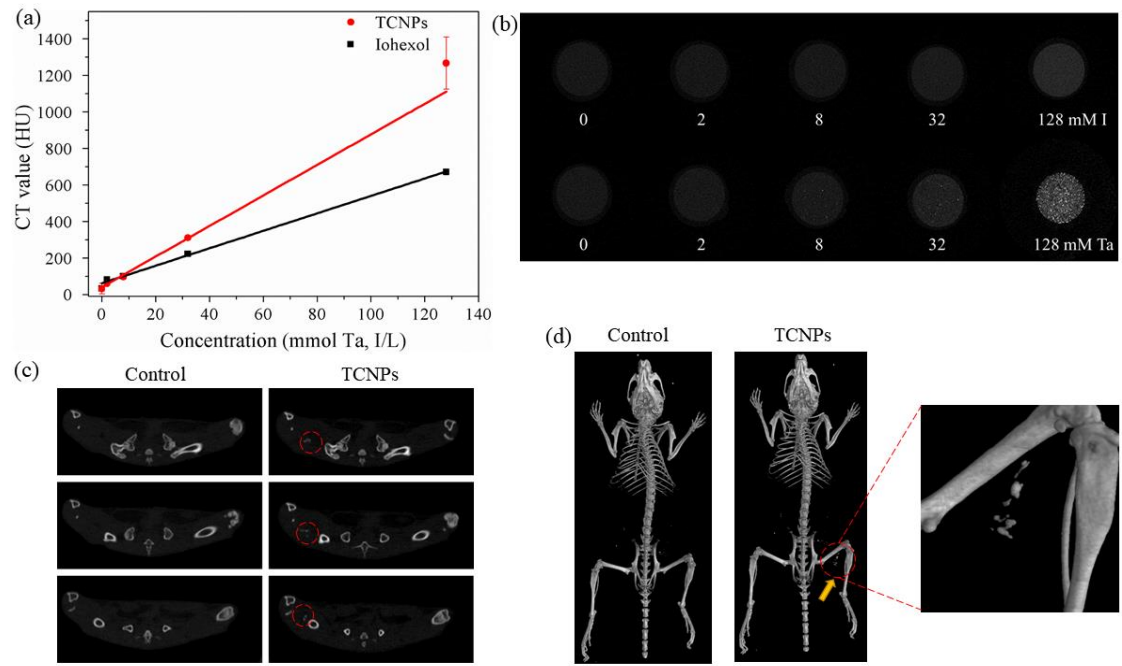

**Figure 4**

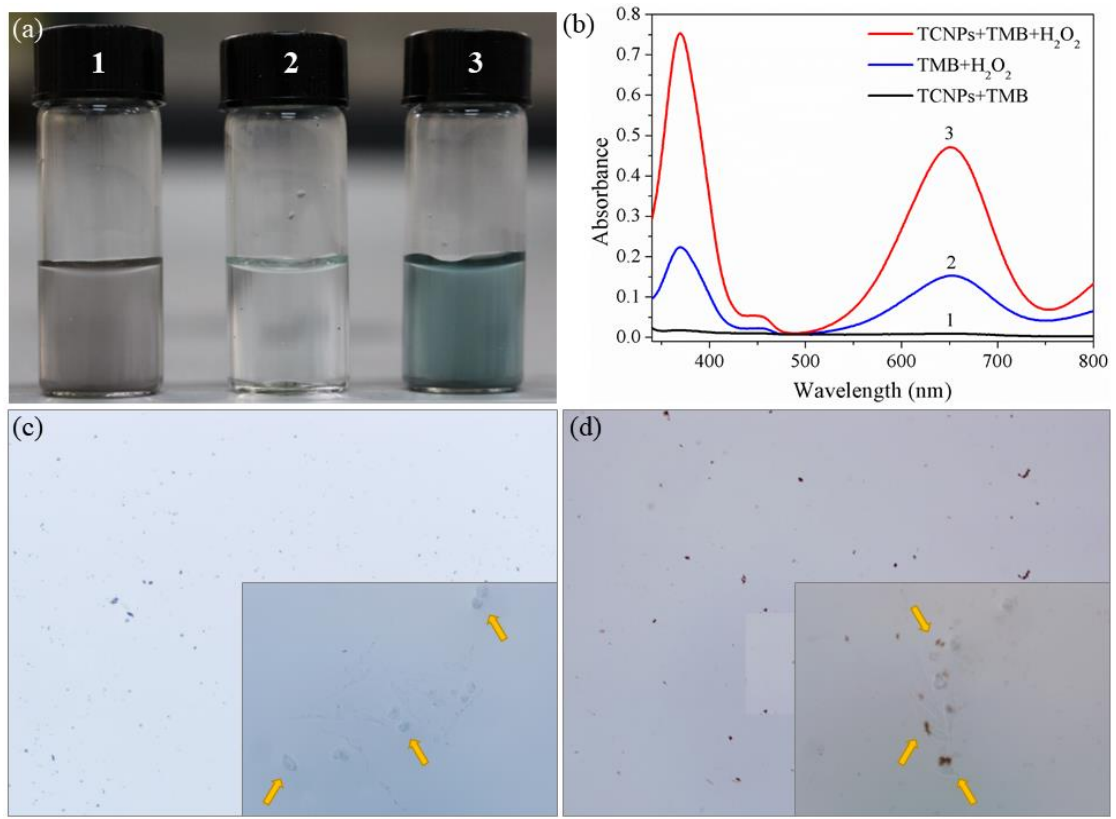

**Figure 5**

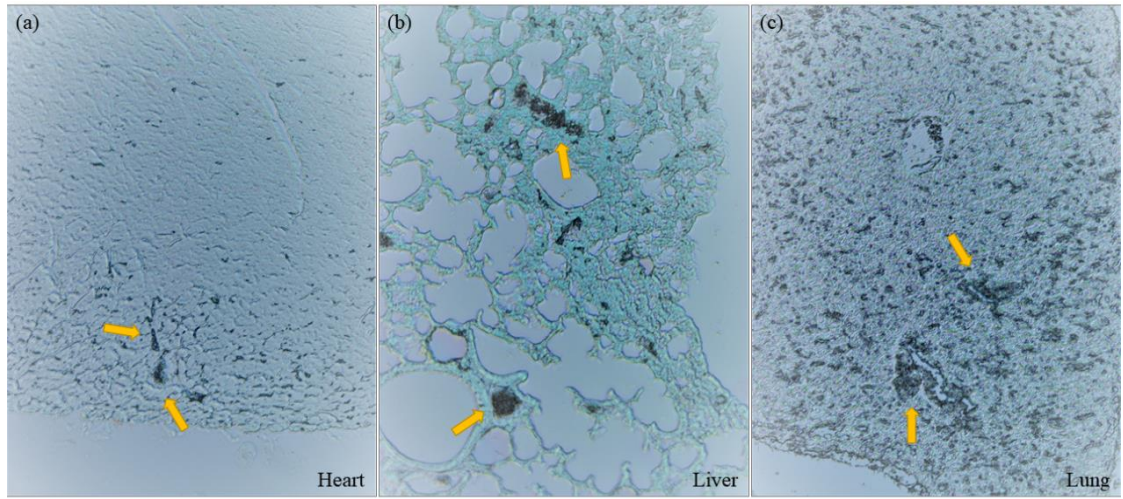